\documentclass[preprint]{revtex4}  % Hochlademodus
\usepackage{amsmath}
\usepackage{amssymb}
\usepackage{amsbsy}
\usepackage{graphicx}
\usepackage{wasysym}
\usepackage[dvips]{epsfig}
\usepackage{psfrag}
\usepackage[ansinew]{inputenc}
\usepackage[T1]{fontenc}
\usepackage{lmodern}
\usepackage{color}

\newcommand{\ul}{\underline}

\newcommand{\p}{\partial}

\newcommand{\bra}[1]{\left\langle{#1}\right|}

\newcommand{\braket}[2]{\left\langle{#1} \,|\, {#2} \right\rangle}

\newcommand{\op}[1]{\mathsf #1}
\newcommand{\set}[1]{\mathcal{#1}}

\renewcommand{\vec}[1]{\mbox{\boldmath $ #1 $}}

\def\gu{\;{\lower0.3ex\hbox{$\buildrel > \over {\scriptstyle \sim}$}}\;}
\def\lu{\;{\lower0.3ex\hbox{$\buildrel < \over {\scriptstyle \sim}$}}\;}

\def\XXint#1#2#3{{\setbox0=\hbox{$#1{#2#3}{\int}$}
     \vcenter{\hbox{$#2#3$}}\kern-.5\wd0}}

\begin{document}

\title{Influence of kinetic effects on the spectrum\\ of a parallel electrode probe}
\author{J. Oberrath}
\affiliation{Leuphana University L\"uneburg,\\Institute of Product and Process Innovation,\\ Volgershall 1, 21339 L\"uneburg, Germany}
%\author{M. Lapke}
%\author{T. Mussenbrock}
\author{R.P. Brinkmann}
\affiliation{Institute of Theoretical Electrical Engineering,\\  
             Department of Electrical Engineering and Information Technologies,\\ Ruhr University Bochum, D-44780 Bochum, Germany}
\date{\today}
\begin{abstract}

\textit{Active Plasma Resonance Spectroscopy} (APRS) denotes a class of diagnostic techniques which utilize the natural ability of plasmas to resonate on or near the electron plasma frequency $\omega_{\rm pe}$. One particular class of APRS can be described in an abstract notation based on functional analytic methods in electrostatic approximation. These methods allow for a general solution of the kinetic model in arbitrary geometry. This solution is given as the response function of the probe-plasma system and is defined by the resolvent of an appropriate dynamical operator. The general response predicts an additional damping due to kinetic effects. This manuscript provides the derivation of an explicit response function of the kinetic APRS model in a simple geometry. Therefore, the resolvent is determined by its matrix representation based on an expansion in orthogonal basis functions. This allows to compute an approximated response function. The resulting spectra show clearly a stronger damping due to kinetic effects.

\end{abstract}
\maketitle

\section{Introduction}

During the last two decades \textit{Active Plasma Resonance Spectroscopy} (APRS) has found renewed interest in the context of industrial applicable measurement devices. APRS denotes a diagnostic technique based on the natural ability of plasmas to resonate on or near the electron plasma frequency $\omega_{\rm pe}$: An electric probe couples a radio frequent signal in the GHz range into a plasma. The spectral response is recorded %(with the same\linebreak  or with another probe) 
and a mathematical model is used to determine plasma parameters like the electron density or electron temperature.   

One particular class of APRS can be described with a model in electrostatic approximation  \cite{takayama1960, messiaen1966, waletzko1967, vernet1975, sugai1999, blackwell2005, scharwitz2009, lapke2008, schulz2015}. The corresponding probes excite surface wave modes which vanish at zero plasma density. Many researchers have made attempts at this task \cite{fejer1964, harp1964, crawford1964, dote1965, kostelnicek1968, cohen1971, tarstrup1972,
aso1973, bantin1974, booth2005, walker2006, lapke2007, xu2009,
xu2010, li2010, liang2011, ichikawa1962, buckley1966, balmain1966, hellberg1968, li1970,
meyer1975, nakatani1976}, using analytical and/or numerical techniques and plasma models of different complexity.  

The cited publications have in common that they all concentrate on specific probe designs. At the Ruhr University Bochum, for example, the \textit{Multipole Resonance Probe (MRP)}, an optimized variant of APRS in electrostatic approximation was invented, analyzed, and characterized \cite{lapke2008, lapke2011, styrnoll2013, styrnoll2014}. However, it is also of interest to study generic features of APRS which are independent of any particular design. Using methods of functional analysis, such an abstract study of APRS based on the cold plasma model is given in \cite{lapke2013}. The main result of that investigation was that, for any possible probe design, the spectral response function could be expressed as a matrix element of the resolvent of the dynamical operator. The correctness and usability of this result was demonstrate in spherical geometry for the impedance probe and the MRP \cite{oberrath2014b}. 

In a recent published paper \cite{oberrath2014}, the authors presented a fully kinetic generalization of the study of \cite{lapke2013}, i.e., an abstract kinetic model of APRS in electrostatic approximation valid for all pressures. It was shown that many insights could be directly transferred. In particular, it still holds that, for any  possible probe design, the spectral response of the probe-plasma system can be expressed as a  matrix element of the resolvent of the dynamical operator. Furthermore, it was shown that the resonances of an APRS probe exhibit a residual damping in the limit of vanishing pressure, which cannot be explained by Ohmic dissipation but only by kinetic effects.

\pagebreak

However, the predicted influence of kinetic effects on the damping is not demonstrated for a spectrum of a probe in an explicit geometry, yet. Within this manuscript we show this influence on the spectrum of a parallel electrode probe. Such a probe is not used in real measurements, but serves as a toy model with the simplest available geometry. To show the kinetic influence, we determine its response function following the same solution strategy for a functional analytic approach as proposed in \cite{oberrath2014b}. Therefore, the response function is expanded by a complete set of orthogonal functions. Truncating this expansion yields an approximated response, which clearly shows additional damping due to kinetic effects.

\section{Model of the Parallel Electrode Probe}\label{sec:model}

In a recent published paper an abstract linearized kinetic model for APRS in electrostatic approximation was derived \cite{oberrath2014}. Here, we apply the results of the abstract model to a certain geometry. It is chosen to be as simple as possible to clearly demonstrate the following algorithm and the influence of kinetic effects on the damping behavior in APRS.  

\begin{figure}[h!]
\psfrag{electrodes}{{\small electrode $\set E_{1/2}$}}
\psfrag{plasma}{{\small plasma ($\set P$)}}
%\psfrag{sheath (S)}{{\tiny sheath $\set S$}}
%\psfrag{dielectric (D)}{\hspace{-.5ex}{\small dielectric ($\set D$)}}
%\psfrag{isolator (I)}{{\small insulator ($\set I$)}}
%\psfrag{electrodes (E)}{{\small electrode (${\set E}_n$)}}
%\psfrag{holder}{{\small holder}}
%\psfrag{shielded cable}{{\small shielded cable}}
%\psfrag{signal generation/}{{\small signal generation/}}
%\psfrag{evaluation}{{\small evaluation}}
%\psfrag{chamber wall}{{\small chamber wall}}
\psfrag{interface}{\hspace{-2ex}{\small interface ($\set F$)}}
%\psfrag{boundary}{\hspace{-2.5ex}{\small interface ($\set F$)}}
\epsfig{file=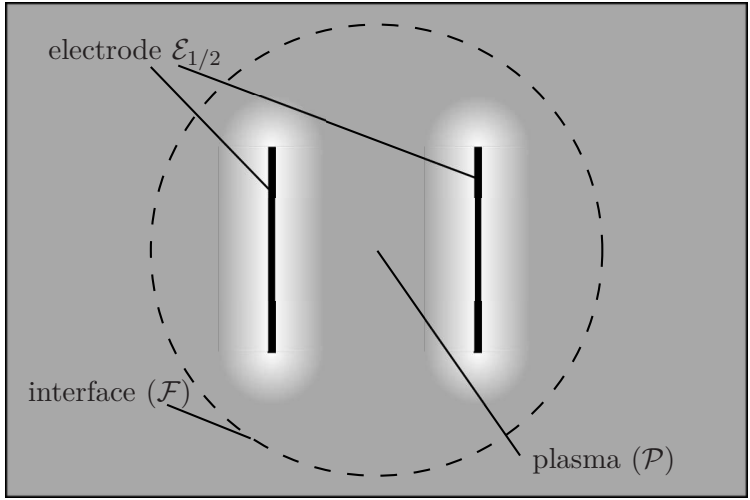,width=0.6\columnwidth}
\caption{Illustration of the idealized model of the PEP. At the electrodes $\set E_{1/2}$ the radio frequent voltages $U_{1/2}(\omega)$ are applied, respectively. The probe perturbs the stationary plasma and the boarder between the perturbed and unperturbed region is called interface $\set F$. }
\label{abstractmodel}
\end{figure}

As depicted in fig.~\ref{abstractmodel} we assume an idealized probe that consists of two plane parallel electrodes with a distance $L$. At the electrodes $\set E_{1/2}$ located by $z=0$ and $z=L$ the radio frequent voltages $U_{1/2}(\omega)$ are applied, respectively, where the frequency $\omega$ is smaller than the electron plasma frequency $\omega_p$. This probe is placed into a stationary plasma and perturbs it. The interface $\set F$ separates the perturbed and the unperturbed region of the plasma. As a further simplification we just focus on the perturbed plasma $\set P$ in between the electrodes. Such a probe is not used in real measurements. It is meant as a toy model and is called \textit{parallel electrode probe (PEP)}. 

The dynamical behavior of the probe-plasma system can be described by the linearized and normalized Boltzmann equation in electrostatic approximation. In this simple geometry the 6 dimensional distribution function reduces to 3 dimensions due to symmetry. It depends on the distance between the electrodes $z\in[0,L]$, the absolute value of the velocity $v\in[0,\infty)$, and the projection angle $\chi\in[0,\pi]$ of $\vec v$ to the $z$ direction and is given by
\begin{align} 
\frac{\p g}{\p t}
&+ v\cos(\chi)\frac{\p g}{\p z} 
 + \frac{\p \bar\Phi}{\p z}
   \left(\cos(\chi)\frac{\p g}{\p v}
   -\frac{\sin(\chi)}{v}\frac{\p g}{\p \chi}\right)
 -v\cos(\chi)\frac{\p \op\Phi}{\p z}
 \nonumber\\
&-\sum_{n=1}^2 U_n v\cos(\chi)\frac{\p \psi_n}{\p z} 
 =\frac{\nu_0}{2}\int_0^{\pi} g \sin(\chi)\, d\chi -\nu_0  g \ .
\label{LinearizedBoltzmannEquationInfinite} 
\end{align}
%\begin{align} 
%\frac{\p g}{\p t}
%&+ \vec{v}\cdot\nabla g + \frac{e}{m}\n\bar\Phi\cdot\n_{v}g-e\vec{v}\cdot\n{\op\Phi}
% \nonumber\\
%&-\sum_{n=1}^N U_n e\vec v\cdot\n\psi_{\rm n}
% =\int_\Omega\frac{d\nu}{d\Omega}g(|\vec v|\vec e)\, d\Omega -\nu_0  g \ .
%\label{LinearizedBoltzmannEquationInfinite} 
%\end{align}
\pagebreak

The perturbed distribution function $g$ of the electrons is defined in $\set P$ with homogeneous boundary conditions at the electrodes $g(0,v,\chi , t)=g(L,v,\chi ,t)$. Pure elastic collisions with a constant collision frequency $\nu_0$ are taken into account between electrons and the neutral background. 
 
$\op\Phi$ is called inner potential. It is a linear functional of $g$ and obeys Poisson's equation with homogeneous boundary conditions
\begin{equation}
\frac{\p^2\op\Phi}{\p z^2} 
= \int_{0}^{\pi} \int_{0}^{\infty} w\, g \sin(\chi)\,dv\, d\chi \ .\label{InPoisson}
\end{equation}
Here, $w$ is a positive weighting function. It is defined as the negative derivative of the equilibrium distribution $F$ with respect to the total energy $\epsilon=\frac{1}{2}v^2-\bar\Phi$, where $\bar\Phi$ is the equilibrium potential. Assuming a Maxwellian distribution $w$ is given by
\begin{equation}
w(z,v)=\frac{1}{\sqrt{2\pi}}e^{-\frac{v^2}{2}+\bar\Phi(z)}
=\frac{n(z)}{\sqrt{2\pi}}e^{-\frac{v^2}{2}}\ .
\end{equation} 

The radio frequent excitation of the probe is represented by the electrode functions $\psi_n$. They follow Laplace's equation
\begin{equation}
\frac{\p^2\psi_n}{\p z^2} = 0\label{Laplace}
\end{equation}
and fulfill the boundary conditions $\psi_n(0,t)=\delta_{1n}$ and $\psi_n(L,t)=\delta_{2n}$ ($\delta_{1n}$ and $\delta_{2n}$ are Kronecker deltas.). Their solutions are easily determined as
\begin{equation}
\psi_1(z)=1-\frac{z}{L}\quad \text{and}\quad \psi_2(z)=\frac{z}{L}\label{charPEP}\ .
\end{equation}

\section{Probe response in functional analytic description}

According to the abstract model presented in \cite{oberrath2014}, the current at one electrode of the PEP can be written in a Hilbert space notation as 
\begin{equation}
i_1=\sum_{n=1}^2\bra{e_1}\left(i\omega-\op T_V-\op T_S\right)^{-1} e_{n}\rangle U_n
   =\sum_{n=1}^2 Y_{1n} U_n\ .
\label{current}
\end{equation}
The admittance between two electrodes is represented by the scalar product between two excitation vectors $e_n=v\cos(\chi)\frac{\p}{\p z}\psi_n$ and the resolvent, which consists of the Vlasov operator $\op T_V$ and the collision operator $\op T_S$,
\begin{equation}
Y_{1n}=\bra{e_1}\left(i\omega-\op T_V-\op T_S\right)^{-1} e_{n}\rangle\ .
\label{response}
\end{equation}
These operators, applied to a general state vector $g$, are defined by
\begin{eqnarray} 
\op T_V g & = & -v\cos(\chi)\frac{\p g}{\p z} 
                -\frac{\p \bar\Phi}{\p z}
                 \left(\cos(\chi)\frac{\p g}{\p v}
                -\frac{\sin(\chi)}{v}\frac{\p g}{\p \chi}\right)
                +v\cos(\chi)\frac{\p \op\Phi}{\p z}\ ,\label{OpTv}\\[1ex]
\op T_S g & = & \frac{\nu_0}{2}\int_0^{\pi} g \sin(\chi)\, d\chi -\nu_0  g \ .
\label{OpStoss} 
\end{eqnarray}
The corresponding scalar product is motivated by the free energy of the probe-plasma system and reduces in the geometry of the PEP to
\begin{eqnarray}
\braket{g'}{g}
& = & \braket{g'}{g}_{\set P}+\braket{g'}{g}_{\set V}\label{ScalarProduct}\\[1ex]
& = &  \int_0^{L}\int_{0}^\pi\int_0^\infty
       {g'}^*\, w\, g\, v^2 \sin(\chi)\,dv \,d\chi \,dz
       %\nonumber\\[1ex]
%&   & 
+\int_{0}^{L}\frac{\partial{\mathsf\Phi}^{\prime*}}{\partial z}
  \frac{\partial{\mathsf\Phi}}{\partial z}\,dz\nonumber \ .%\\[1ex]
%& = & 
%\int_0^{2\pi}\int_{-1}^1\int_R^{R_\infty} \int_0^{2\pi}\int_{-1}^1\int_0^\infty
% {g'}^* w g r^2 \sqrt{2\varpi} \,d\varpi\,d\psi\,d\xi\,dr\,d\zeta\,d\vp
% \nonumber\\[1ex] 
%&   & 
%-\int_0^{2\pi}\int_{-1}^1\int_R^{R_\infty} \int_0^{2\pi}\int_{-1}^1\int_0^\infty
% \epsi_r {\op\Phi'}^* w g r^2 \sqrt{2\varpi} 
% \,d\varpi\,d\psi\,d\xi\,dr\,d\zeta\,d\vp\nonumber \ .
\end{eqnarray}
 
Following the functional analytic approach, we have to expand the admittance $Y_{1n}$ by means of a complete orthonormal basis $\{l\}$ of the Hilbert space. Introducing the corresponding completeness relation twice into equation \eqref{response} yields
\begin{equation}
Y_{1n}%=\bra{e_n}\sum_{k'}\ket{z_{k'}}\bra{z_{k'}}
%\frac{1}{i\omega\op{I}-\op{T}_C-\op{T}_D} \sum_{k}\ket{z_k}\braket{z_{k}}{e_{n'}}
=\sum_{l'}\braket{e_1}{l'}\sum_{l}\langle l' |
\left(i\omega-\op{T}_V-\op{T}_S\right)^{-1} l\rangle\braket{l}{e_n}\ .
\label{responseExp}
\end{equation}
This is a vector-matrix-vector product which is determined by the algebraic representation of the resolvent. In \cite{oberrath2014b} it is shown that the matrix representation of a resolvent $(i\omega-\op T)^{-1}$ is given by the inverse of the matrix representation of the operator $i\omega-\op T$. It allows first to determine the matrix representation of the operator and than to compute the inverse to get the matrix representation of the resolvent. 

Now, the strategy to determine the admittance in a certain geometry is obvious. We choose an appropriate orthonormal basis, determine the matrix representation of $i\omega-\op T_V-\op T_S$, calculate its inverse to get the matrix representation of the resolvent, compute the scalar products $\braket{l}{e_n}$ and $\braket{e_n}{l'}$, to finally get the admittance as a vector-matrix-vector multiplication.

\section{Complete orthogonal basis in a simplified geometry}\label{sec:geometry}

%To reduce the model to the pure mathematical structure it can be normalized. All distances are measured in Debye length $\lambda_{\rm D}$ and the velocity in the thermal velocity of the electrons $v_{\rm t}$. The collision frequency is normalized to the plasma frequency $\omega_{\rm p}$ and the time to $\omega_{\rm p}^{-1}$. All potentials are measured in $\frac{T}{e}$, where $T$ is the electron temperature and $e$ the elementary charge. The perturbed distribution function is normalized to $T$ and the electron density to $n_0$, the maximum value at $z=\frac{L}{2}$. 

%The normalized scalar product in this geometry can be simplified
%\begin{align}
%\langle g^\prime | g \rangle 
%= \int_{0}^L\int_{0}^{\infty}\int_{0}^{\pi}&  g^{\prime*} g\, w \, v^2\sin(\chi)\,d\chi\,dv\, dz\nonumber\\[1ex]
%&+\int_{0}^{L}\varepsilon_0\frac{\partial{\mathsf\Phi}^{\prime*}}{\partial z}
%  \frac{\partial{\mathsf\Phi}}{\partial z}\,dz\ . 
%\label{ScalarProductPEP}
%\end{align}
Before the admittance can explicitly be expanded, we have to find an appropriate set of basis functions. These basis functions have to fulfill the boundary conditions and should be orthogonal in the scalar product \eqref{ScalarProduct}. The first part of the scalar product depends on the weighting function $w$, which has a crucial influence on the choice of orthogonal basis functions. Due to the exponential part the generalized Laguerre polynomials $L_\kappa^{\lambda+\frac{1}{2}}\left(\frac{1}{2}v^2\right)$ are an adequate choice for a complete basis in the absolute value of the velocity. They become the orthonormal functions $\Lambda_{\kappa}^{\lambda}(v)$ %\cite{balescu1988} 
with an additional factor  
\begin{equation}
\Lambda_{\kappa}^{\lambda}(v)
=\sqrt{\frac{\kappa !}{\Gamma(\kappa+\lambda+\frac{3}{2})}}
 \left(\frac{v^2}{2}\right)^{\frac{\lambda}{2}} 
      L_\kappa^{\lambda+\frac{1}{2}}\left(\frac{v^2}{2}\right)\ .%\ ,\\[1ex]
%\tilde P_{k}^l(r)
%& = & \left(\frac{R}{r}\right)^3 
%      e^{-\frac{1}{2}\bar\Phi(r)-iK\frac{R}{r}} j_{l}\left(K_l^k \frac{R}{r}\right) \ .     
\end{equation}
On the interval $\chi\in[0,\pi]$ Legendre polynomials $P_\lambda(\cos(\chi))$ yield also a complete orthonormal set with an additional factor
\begin{equation}
\bar P_\lambda(\chi)
=\sqrt{\frac{2\lambda+1}{2}}P_\lambda(\cos(\chi))\ .
\end{equation}
$\lambda\in\mathbb N_0$ is the expansion index for the projection angle and $\kappa\in\mathbb N_0$ for the absolute value of the velocity.

In physical space it is difficult to determine an orthogonal function due to the two different parts of the scalar product. %It can be found numerically, but we follow a different 
Therefore, we assume $\tilde P_k(z)$ as complete basis function on the interval $[0,L]$ which fulfills the boundary conditions $\tilde P_k(0)=\tilde P_k(L)=0$. Its distribution will remain undefined, yet. Together with the basis functions $\Lambda_{\kappa}^\lambda(v)$ and $\bar P_\lambda(\chi)$ we can define the following complete basis vector, which is orthonormal on the reduced velocity space, 
\begin{equation}
g_{k}^{\kappa\lambda}(z,v,\chi)
=\tilde P_k(z)\Lambda_{\kappa}^\lambda(v)\bar P_\lambda(\chi)\label{BasisKet}\ .
\end{equation}
%The dependencies of the basis functions in velocity space propose the substitutions $\xi=\cos(\chi)$ and $\varpi=\frac{1}{2}v^2$ which simplify the scalar product  
%\begin{eqnarray}
%\braket{g'}{g}
%& = & \braket{g'}{g}_{\set P}+\braket{g'}{g}_{\set V}\\[1ex]
%& = &  \int_0^{L}\int_{-1}^1\int_0^\infty
%       {g'}^* w g \sqrt{2\varpi} \,d\varpi\,d\xi \,dz
%       \nonumber\\[1ex]
%&   & +\int_{0}^{L}\frac{\partial{\mathsf\Phi}^{\prime*}}{\partial z}
%  \frac{\partial{\mathsf\Phi}}{\partial z}\,dz\nonumber \ .%\\[1ex]
%& = & 
%\int_0^{2\pi}\int_{-1}^1\int_R^{R_\infty} \int_0^{2\pi}\int_{-1}^1\int_0^\infty
% {g'}^* w g r^2 \sqrt{2\varpi} \,d\varpi\,d\psi\,d\xi\,dr\,d\zeta\,d\vp
% \nonumber\\[1ex] 
%&   & 
%-\int_0^{2\pi}\int_{-1}^1\int_R^{R_\infty} \int_0^{2\pi}\int_{-1}^1\int_0^\infty
% \epsi_r {\op\Phi'}^* w g r^2 \sqrt{2\varpi} 
% \,d\varpi\,d\psi\,d\xi\,dr\,d\zeta\,d\vp\nonumber \ .
%\end{eqnarray}

As constraint the basis functions have to fulfill Poisson's equation \eqref{InPoisson}. Entering \eqref{BasisKet} into \eqref{InPoisson}, it can be simplified by means of the orthogonality relations of the Legendre and Laguerre polynomials
\begin{equation}
\frac{\p^2\op\Phi_{k}}{\p z^2}
%& = &
%\frac{\tilde P_k(z)n(z)}{(2\pi)^{\frac{3}{2}}}
%\int_0^{2\pi}\int_{-1}^1\int_0^{\infty}\sqrt{2\varpi} e^{-\varpi}
%\Lambda_{\kappa}^\lambda P_\lambda^\mu \,e^{i\mu\psi}\,d\varpi\,d\xi\,d\psi \\[1ex]
%& = &
%\frac{\tilde P_k(z)n(z)}{\sqrt{\pi}}
%\int_{-1}^1\int_0^{\infty}
%\sqrt{\varpi} e^{-\varpi}\Lambda_{\kappa}^\lambda P_\lambda 
%\,d\varpi\,d\xi \nonumber
%\\[1ex]
%& = &
%\frac{2\delta_{\lambda,0}}{\sqrt{\pi}}\tilde P_k(z)n(z)
%\int_0^{\infty}\sqrt{\varpi} e^{-\varpi} L_{\kappa}^{\frac{1}{2}}\,d\varpi \nonumber
%\\[1ex]
%& = & 
%\frac{2\Gamma\left(\frac{3}{2}\right)}{\sqrt{\pi}} 
%\,\delta_{\lambda,0}\,\delta_{\kappa,0}\,\tilde P_k(z)n(z)
%\\[1ex]
%& = &
=\frac{\delta_{\lambda 0}\,\delta_{\kappa 0}}{\pi^{\frac{1}{4}}}\,\tilde P_k(z)n(z)\label{PoissonPEP}\ .
\end{equation}
With the boundary conditions $\op\Phi_{k}(0)=\op\Phi_{k}(L)=0$ this differential equation can be solved by integration
\begin{align}
\op\Phi_{k}(z)
=&\frac{1}{\pi^{\frac{1}{4}}}\int_0^z \int_0^{z'} \tilde P_k(z'')n(z'') \, dz''\, dz'
  \,\delta _{\kappa 0}\, \delta_{\lambda 0}\\[1ex]
 &-\frac{z}{\pi^{\frac{1}{4}}L}\int_0^L \int_0^{z'} \tilde P_k(z'') n(z'') \, dz'' \, dz'
  \,\delta _{\kappa 0}\, \delta_{\lambda 0}\nonumber \ .
\end{align}
In the scalar product \eqref{ScalarProduct} we need just the derivative of the potential which is given by
\begin{align}
\frac{\p\op\Phi_{k}}{\p z}
=&\frac{1}{\pi^{\frac{1}{4}}}
  \int_0^z \tilde P_k(z')n(z')\, dz'\,\delta _{\kappa 0}\, \delta_{\lambda 0}\\[1ex]
 &-\frac{1}{\pi^{\frac{1}{4}}L}\int_0^L \int_0^{z'} \tilde P_k(z'') n(z'') \, dz'' \, dz'
  \,\delta _{\kappa 0}\, \delta_{\lambda 0} \nonumber\ .
\end{align}

%\begin{gather}
%\begin{array}{rclcrclcrclcrcl}
%          \vec r & \ra & \lambda_D \tilde{\vec r}                & , & 
%          \vec v & \ra & v_{\rm t} \tilde{\vec v}               & , &      
%       n_{\rm i} & \ra & n_0\tilde{n}_{\rm i}                    & , &    
%             g & \ra & T \tilde{g}                               \ ,\\[2ex]          
%        \bar\Phi & \ra & \ds\frac{T}{e}\,\tilde{\bar\Phi}          & , &   
%            \phi & \ra & \ds\frac{T}{e}\,\tilde{\phi}              & , & 
%\phi^{\rm (vac)} & \ra & \ds\frac{T}{e}\,\tilde{\phi}^{\rm (vac)}  & , & 
%               F & \ra & \ds\frac{n_0}{v_{\rm t}^3}\tilde{F}    \ ,\\[2ex]
%               t &\ra &\omega_{\rm pe}^{-1} \tilde{t}            & , & 
%             \nu & \ra & \omega_{\rm pe} \tilde{\nu}             & . & &&&    & 
%\end{array}
%\end{gather}

%\begin{equation}
%\sum_{k,l,m}\sum_{\kappa ,\lambda ,\mu} %\ket{z_{klm}^{\kappa\lambda\mu}}\bra{z_{klm}^{\kappa\lambda\mu}}=1
%\label{approxVollRel}
%\end{equation}\\[-5mm]

%\begin{eqnarray}
%\op T_V\ket z
%&=&-\frac{\p\Phi}{\p z}\left(
%    \frac{\left(1-\xi ^2\right)}{\sqrt{2\varpi}} \frac{\p g}{\p\xi}
% +  \sqrt{2\varpi}\xi \frac{\p g}{\p\varpi}\right)\nonumber\\[1ex]
%& & -\sqrt{2\varpi}\xi \left(\frac{\p g}{\p z}-\frac{\p \op\Phi}{\p z}\right)\ ,
%     \\[1ex]%\ .
%\op T_S\ket z & = & \frac{\nu}{2}\int_{-1}^1 g(\varpi,\xi)\,d\xi -\nu g%\label{TsPEP}
%\end{eqnarray}

\section{Basis matrix}\label{sec:basis}

In the previous section we introduced a complete set of basis vectors $g_{k}^{\kappa\lambda}$. Following the solution strategy, the scalar product of two basis vectors has to be determined. They are orthonormal in the velocity space which allows to simplify the first part of the scalar product by means of the corresponding orthogonality relations
\begin{equation}
\braket{g_{k'}^{\kappa' \lambda'}}{g_{k}^{\kappa\lambda}}_{\set P}
=\delta_{\kappa\kappa'}\, \delta_{\lambda\lambda'}\, \op B_{kk'}^{(\kappa\lambda)}
\end{equation}
with
\begin{equation}
\op B_{kk'}^{(\kappa\lambda)}=
\frac{1}{\sqrt{\pi}}\int_0^L n \tilde P_{k'}^* \tilde P_k\, dz\ .
\end{equation}
The second part of the scalar product is determined by the derivative of the potentials $\op\Phi_{k}$ and $\op\Phi_{k'}^*$. Due to the corresponding Kronecker Deltas it reduces with $\kappa=\kappa'=\lambda=\lambda'=0$ to
\begin{equation}
 \braket{g_{k'}^{0 0}}{g_{k}^{00}}_{\set V}
=\int_0^L \frac{\p{\op\Phi_{k'}}^*}{\p z}
          \frac{\p\op\Phi_{k}}{\p z}\,dz\ .
\end{equation}
For all other combinations of the expansion indices in the velocity space $\braket{g_{k'}^{\kappa' \lambda'}}{g_{k}^{\kappa\lambda}}_{\set V}=0$. Owing to that we split up the complete scalar product into
\begin{eqnarray}
\braket{g_{k'}^{00}}{g_k^{00}}
& = &  \op B_{kk'}^{(00)}
      +\int_0^L \frac{\p{\op\Phi_{k'}}^*}{\p z}
       \frac{\p\op\Phi_{k}}{\p z}\,dz=\op B_{kk'}^{(0)}\ ,\label{BasisSP0}      \\[1ex]
%& = &  
\braket{g_{k'}^{\kappa' \lambda'}}{g_{k}^{\kappa\lambda}}
& = &  \delta_{\kappa\kappa'}\,\delta_{\lambda\lambda'}\, \op B_{kk'}^{(\kappa\lambda)}
       \, . 
\end{eqnarray}

One can see, that the part $\braket{g'}{g}_{\set V}$ of the complete scalar product, which made it difficult to find an orthogonal set of basis functions in the physical space, is only given in \eqref{BasisSP0}. Due to the separation of the integrals in physical and velocity space we are able to first determine the matrices $\ul{\op B}_{\, k}^{(0)}$ and $\ul{\op B}_{\, k}^{(\kappa\lambda)}$ for the indices $k$ and $k'$ with a non-orthogonal basis. Then these matrices can be diagonalized with the rotation matrices $\ul{\op C}^{(0)}$ and $\ul{\op C}^{(\kappa\lambda)}$ which means a change of the basis to an orthogonal one. The diagonalized matrices read as follows
\begin{eqnarray}
\ul{\op D}_{\, k}^{(0)}
& = &  \ul{\op C}^{(0)}\,\ul{\op B}_{\, k}^{(0)}\, {\ul{\op C}^{(0)}}^T\ , 
       \label{BasisD0}\\[1ex]
\ul{\op D}_{\, k}^{(\kappa\lambda)}
& = &  \delta_{\kappa\kappa'}\, \delta_{\lambda\lambda'}
       \ul{\op C}^{(\kappa\lambda)}\,\ul{\op B}_{\, k}^{(\kappa\lambda)}\, 
       {\ul{\op C}^{(\kappa\lambda)}}^T\hspace{-1.5ex}.\hspace{1.5ex}\quad 
\label{BasisDkl}       
\end{eqnarray}
After that we multiply these diagonal matrices with their inverse to find the corresponding identity matrix. Finally we arrive at the basis identity matrix $\ul{\op I}$ of the complete scalar product with the indices of the velocity space. %This identity matrix could have been also derived directly with an orthonormal basis in the physical space. 

\section{Matrix of the Vlasov and the collision operator}\label{sec:TvTs}

After the basis matrix we want to determine the matrix of the Vlasov operator. To evaluate its matrix elements we need the derivative of the inner potential $\op\Phi^{(\op T_V)}$ caused by the vector $\op T_V g$. Entering this vector in Poisson's equation one can find, that the derivative of $\op\Phi^{(\op T_V)}$ is given by the electron particle flux. In the geometry of the PEP, we find for a basis vector
\begin{equation}
\frac{\p\op\Phi_k^{(\op T_V)}}{\p z}
=-n\tilde P_k \,\int_{0}^\pi \int_0^{\infty }\frac{e^{-\frac{1}{2}v^2}}{\sqrt{2\pi}}  
    \Lambda_\kappa^{\lambda} \bar P_\lambda
    \cos(\chi)\sin(\chi)v^3\,d v \,d\chi \ .
\end{equation}
Due to that the explicit solution of the potential $\op\Phi^{(\op T_V)}$ is not needed and the scalar product of two basis vectors and $\op T_V$ is determined by
\begin{align}
\langle g_{k'}^{\kappa'\lambda'} | \op T_V g_{k}^{\kappa\lambda}\rangle
= & \ {\op V}_{kk'}^{(1)}
    \int_{0}^\pi \int_0^{\infty } 
    \frac{e^{-\frac{1}{2}v^2}}{\sqrt{2\pi}}\Lambda_{\kappa'}^{\lambda'} \bar P_{\lambda'} 
    \left(\frac{\sin(\chi)}{v}\Lambda_\kappa^{\lambda} 
    \frac{\p \bar P_\lambda}{\p\chi}
   -\cos(\chi) \bar P_\lambda\frac{\p\Lambda_\kappa^{\lambda}}{\p v}\right)\sin(\chi)v^2
    \,d v \,d\chi\nonumber\\[1ex]
  &-{\op V}_{kk'}^{(2)}
    \,\int_{0}^\pi \int_0^{\infty }\frac{e^{-\frac{1}{2}v^2}}{\sqrt{2\pi}}  
    \Lambda_{\kappa'}^{\lambda'} \bar P_{\lambda'}\Lambda_\kappa^{\lambda} \bar P_\lambda
    \cos(\chi)\sin(\chi)v^3\,d v \,d\chi\nonumber\\[1ex]
  &+{\op V}_{kk'}^{(3)}
    \,\int_{0}^\pi \int_0^{\infty }\frac{e^{-\frac{1}{2}v^2}}{\sqrt{2\pi}}  
    \Lambda_{\kappa'}^{\lambda'} \bar P_{\lambda'}
    \cos(\chi)\sin(\chi)v^3\,d v \,d\chi\nonumber\\[1ex] 
  &-{\op V}_{kk'}^{(4)}
    \,\int_{0}^\pi \int_0^{\infty }\frac{e^{-\frac{1}{2}v^2}}{\sqrt{2\pi}}  
    \Lambda_\kappa^{\lambda} \bar P_\lambda
    \cos(\chi)\sin(\chi)v^3\,d v \,d\chi
\end{align}
with
\begin{eqnarray}
{\op V}_{kk'}^{(1)} 
& = & \int_0^L \tilde P_{k'}^*\tilde P_k \frac{\p n}{\p z}\, dz\ ,\\[1ex]
{\op V}_{kk'}^{(2)} 
& = & \int_0^L \tilde P_{k'}^* n\frac{\p \tilde P_k}{\p z}\, dz\ ,\\[1ex]
{\op V}_{kk'}^{(3)}
& = & \int_0^L \tilde P_{k'}^* n\frac{\p\op\Phi_k^{(\set P)}}{\p z}\, dz\ ,\\[1ex]
{\op V}_{kk'}^{(4)} 
& = & \int_0^L \tilde P_{k} n \frac{\p\op\Phi_{k'}^{(\set P)\,*}}{\p z}\, dz\ .
\end{eqnarray}

The factorization of the integrals allows for the same evaluation as done for the basis matrix. We determine the matrices for the physical space ${\ul{\op V}_{\,k}^{(i)}}$ with a non-orthogonal basis over the indices $k$ and $k'$ with $i=1,2,3,4$. Multiplying them with the rotation matrices and the inverse diagonal matrices of the previous section leads to the inner block matrices for the physical space.

As shown in \cite{oberrath2014} the Vlasov operator is skew self-adjoint. This property transfers to the matrix representation -- called skew hermitian -- if the basis vector is orthogonal. In fact the basis vector is orthonormal in the velocity space and leads to a vanishing scalar product for $\kappa=\kappa'=\lambda=\lambda'=0$. Due to that we only need the rotation matrix $\ul{\op C}^{(\kappa\lambda)}$ and the inverse diagonal matrix ${\ul{\op D}^{(\kappa\lambda)}}^{-1}$ to determine the inner matrices $\ul{\op T}_{Vk}^{(i)}=\ul{\op C}^{(\kappa\lambda)}{\ul{\op V}_{\,k}^{(i)}}\ul{\op C}^{(\kappa\lambda)\,T}{\ul{\op D}^{(\kappa\lambda)}}^{-1}$ for the physical space. Integration in the velocity space (The integrals can analytically be solved, but lead to long expressions.) yields block matrices $\ul{\op T}_V^{(i)}$ and their sum builds the complete matrix of the Vlasov operator 
\begin{equation}
\ul{\op T}_V=\ul{\op T}_V^{(1)}+\ul{\op T}_V^{(2)}+\ul{\op T}_V^{(3)}+\ul{\op T}_V^{(4)}
\label{MTvPEP}\ .
\end{equation}

The last matrix to determine is that of the collision operator. Owing to the fact that an integral in velocity space over the collision operator equals zero, yields also $\braket{g}{\op T_S g}_{\set V}=0$. Therefore, the scalar product of two basis vectors and $\op T_S$ simplifies with the orthogonality relations of the Legendre and Laguerre polynomials to
\begin{align}
\langle g_{k'}^{\kappa'\lambda'} | \op T_S g_{k}^{\kappa\lambda}\rangle
=\langle g_{k'}^{\kappa'\lambda'} | \op T_S g_{k}^{\kappa\lambda}\rangle_{\set P}
%& = & \frac{\nu}{\sqrt{\pi }}\int_0^L \tilde P_{k'} \tilde P_k n\, dz
%      \int_0^{\infty }\int_{-1}^1 e^{-\varpi } \sqrt{\varpi }      
%      \Lambda_{\kappa'}^{\lambda'} P_{\lambda'}\Lambda_\kappa^{\lambda}
%      \left(\frac{1}{2}\int_{-1}^1 P_\lambda \,d\xi- P_\lambda\right) \,d\xi \,d\varpi 
%      \nonumber\\[1ex]
=\, \nu_0
      \left(\delta_{\lambda 0}\,\delta_{0 \lambda'}-\delta_{\lambda\lambda'}\right) 
      \,\delta_{\kappa\kappa'}\, \op S_{kk'}\label{MeTs}
\end{align}
with
\begin{equation}
\op S_{kk'}=\frac{1}{\sqrt{\pi}}\int_0^L n \tilde P_{k'}^* \tilde P_k \, dz
=\op B_{kk'}\label{MeTsRest}\ .
\end{equation}
%\pagebreak

The matrix elements \eqref{MeTs} obviously vanish if $\lambda=\lambda'=0$. For all other elements \eqref{MeTsRest} shows, that they are equal to the elements of the basis matrix. Hence, the multiplication of $\ul{\op S}^{(\kappa\lambda)}$ with the rotation matrix $\ul{\op C}^{(\kappa\lambda)}$ and the inverse matrix ${\ul{\op D}^{(\kappa\lambda)}}^{-1}$ leads to identity matrices in the physical space. By means of the indices of the velocity space we find the final matrix of the collision operator $\ul{\op T}_{\, S}$ as a diagonal matrix multiplied by the negative factor $-\nu_0$. The elements on the main diagonal are zero if $\lambda=\lambda'=0$ otherwise one. 
%\begin{equation}
%\ul{\op T}_{\,Sk}^{(\kappa\lambda)}
%=\frac{2\Gamma\left(\kappa+\lambda+\frac{3}{2}\right)}{\sqrt{\pi}(2\lambda+1)\kappa !}
% \,\delta_{\lambda\lambda'}\,\delta_{\kappa\kappa'}
% \ul{\op C}^{(\kappa\lambda)}\,\ul{\op S}_{\, k}^{(\kappa\lambda)}\, 
% {\ul{\op C}^{(\kappa\lambda)}}^T \label{TsDkl}\ .       
%\end{equation}
%With the indices of the velocity space we find again a diagonal block matrix, where the main diagonal has zero elements if $\lambda=\lambda'=0$. The matrix of the collision operator is given by
%\begin{equation}
%\ul{\op T}_{\, S}=\nu\,\begin{pmatrix}
%\ul 0 & \ul 0 & \dots &&&&& \\[1ex] 
%\ul 0& \ul{\op T}_{\,Sk}^{(01)} & \ul 0 & \dots &&&&\\[1ex]
%\vdots & \ul 0 & \ul{\op T}_{\,Sk}^{(02)} & \ul 0 & \dots &&&\\[1ex]
%& \vdots & \ddots & \ddots & \ddots & \dots &&\\[1ex]
%&& \vdots & \ul 0 & \ul 0 & \ul 0 & \dots &\\[1ex]
%&&& \vdots & \ul 0 & \ul{\op T}_{\,Sk}^{(11)} & \ul 0 & \dots \\[1ex]
%&&&& \vdots & \ul 0 & \ul{\op T}_{\,Sk}^{(12)} & \ddots  \\[1ex]
%&&&&& \vdots & \ddots &\ddots  
%\end{pmatrix}\ .
%\end{equation}
Such a matrix is symmetric and negative semi-definite, like the collision operator itself. Now, the matrix representation of the complete operator $i\omega-\op T_V-\op T_S$ is determined and can be inverted to find the matrix representation of the resolvent.

\section{Excitation vector}

Before we evaluate the expanded admittance \eqref{responseExp}, the scalar products between the basis and the excitation vectors are needed. The latter, with solutions \eqref{charPEP}, are given by
\begin{equation}
e_{1/2}=\mp v\cos(\chi)\frac{1}{L}\label{EVek}\ .
\end{equation}
Entering \eqref{EVek} into the scalar product with the basis vector the scalar products are determined by 
\begin{equation}
 \braket{z_{k'}^{\kappa'\lambda'}}{e_{1/2}}
=\braket{z_{k'}^{\kappa'\lambda'}}{e_{1/2}}_{\set P}
%&=&\sqrt{\frac{2}{\pi}}\int_0^{\infty }\int_{-1}^1\xi e^{-\varpi } \varpi     
%   \Lambda_{\kappa'}^{\lambda'} P_{\lambda'}\,d\xi\,d\varpi\,
%   \int_0^L \tilde P_{k'} n \frac{\p\psi_1}{\p z} \, dz
=\mp\underbrace{\frac{\pi^{\frac{1}{4}}}{2L}\int_0^L \tilde P_{k'}^* n \, dz}_{\vec e_{k'}}
    \,\delta_{\lambda' 1}\,\delta_{\kappa' 0}\ .
\end{equation}
Multiplying $\vec e_{k'}$ from the left side with the rotation matrix $\ul{\op C}^{(\kappa\lambda)}$ and the inverse diagonal matrix ${\ul{\op D}^{(\kappa\lambda)}}^{-\frac{1}{2}}$ yields the excitation vectors $\vec e_{1k'}$ and $\vec e_{2k'}$. (${\ul{\op D}^{(\kappa\lambda)}}^{-\frac{1}{2}}$ is meant as an inverse diagonal matrix, where the square root of the elements on the main diagonal is evaluated.)
%\begin{equation}
%\vec e_{1k'}=\ul{\op C}^{(\kappa\lambda)} \vec e_{k'}\ .
%\end{equation} 
With the expansion indices of the velocity space we find the final excitation vectors $\vec e_{1}=-\vec e_{2}$, which have non-vanishing elements only for $\kappa'=0$ and $\lambda'=1$,
\begin{equation}
\vec e_{1/2}=\left(\vec 0\, ,\, \mp{\ul{\op D}^{(\kappa\lambda)}}^{-\frac{1}{2}} \ul{\op C}^{(01)} \vec e_{k'}\, ,\, \vec 0\, ,\, \dots \right)^T\ .
\end{equation}

\section{Resonance behaviour of the PEP}

Now, we are equipped with all necessary elements to determine the current $i_1$ including the expanded admittance, which is given as a vector-matrix-vector multiplication,
\begin{equation}
i_1=\sum_{n=1}^2\vec e_1^T\cdot
   \left(i\omega \ul{\op I}-\ul{\op T}_{\, V}-\ul{\op T}_{\, S}\right)^{-1}
   \cdot\vec e_n U_n\ .
%\label{current}
\end{equation}
Applying symmetric voltages $U_1=-U_2=\frac{1}{2}U$ at the electrodes simplifies the current
\begin{equation}
i_1=\vec e_1^T\cdot
    \left(i\omega \ul{\op I}-\ul{\op T}_{\, V}-\ul{\op T}_{\, S}\right)^{-1}
    \cdot\vec e_1 U=Y\, U
%\label{current}
\end{equation}
and leads to the complete admittance of the PEP
\begin{eqnarray}
Y
%&=&\sum_{k,k'}\sum_{\kappa,\kappa'}\sum_{\lambda,\lambda'} 
%   \braket{e_1}{z_{k}^{\kappa\lambda}}
%   \left(\bra{z_{k'}^{\kappa'\lambda'}} i\omega+T_V+T_S 
%   \ket{z_{k}^{\kappa\lambda}}\right)^{(-1)}
%   \braket{z_{k'}^{\kappa'\lambda'}}{e_1}\nonumber\\[1ex]
&=&\vec e_1^T\cdot
   \left(i\omega \ul{\op I}-\ul{\op T}_{\, V}-\ul{\op T}_{\, S}\right)^{-1}
   \cdot\vec e_1\ .
\end{eqnarray}
It allows to analyse the spectra of the PEP. To determine approximated spectra we have to evaluate the last integrals in the matrix and vector elements, which are dependent on the basis functions in physical space and the equilibrium electron density $n$. As complete set of basis functions in physical space we choose
\begin{equation}
\tilde P_k(z)=\frac{1}{\sqrt{L}}\sin\left(\frac{k\pi z}{L}\right)\ \text{with}\ k\in\mathbb N\ .
\end{equation}
They fulfill all conditions claimed in section \ref{sec:geometry}. The equilibrium density of the electrons has to have a strong increase and decrease in front of the electrodes and a relatively homogeneous distribution in the bulk. The following combination of exponential functions is a simplification compared to a real density, but it fulfills the important requirements
\begin{equation}
n(z)=\frac{A-e^{-\frac{az}{L}}-e^{\frac{a(z-L)}{L}}}{A-2e^{-\frac{az}{L}}}\ .
\end{equation}
The density is symmetric between the electrodes and normalized to the value in the center. $A$ determines the density value in front of the electrodes and $a$ the strength of the gradient. We chose the simplified density to allow for analytic solutions of all integrals, because large matrices are expected for converged spectra.  

The presented approximated spectra are computed with the fixed parameters $A=1.05$, $a=10$, and $L=100$. They are chosen to represent a reasonable equilibrium density. A typical collision frequency in a low-temperature plasma is about $\nu_0=0.02$. $L$ and $\nu_0$ are normalized to the Debye length $\lambda_D$ and the plasma frequency $\omega_p$, respectively. We use the maximum expansion index $k_{\rm max}=1000$ in physical space to ensure converged spectra for all chosen parameters.
\pagebreak

\begin{figure}[h!]
\centering
\psfrag{Y}{{\tiny $\Re\{Y\}\ [a.u.]$}}
\psfrag{o}{{\small $\frac{\omega}{\omega_p}$}}
\epsfig{file=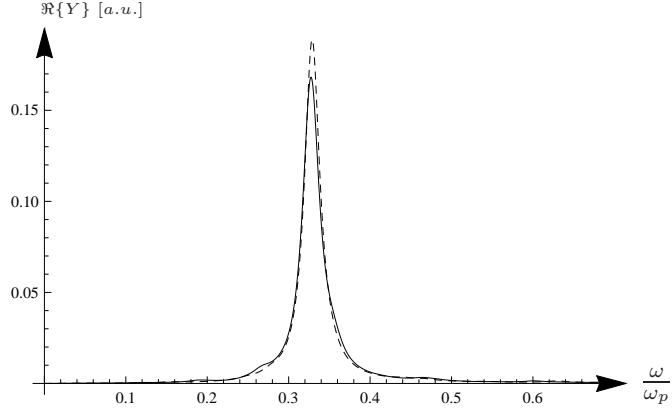,width=8.5cm}
%\hspace{8mm}
%\epsfig{file=Bilder/kinetic/Spielmodell/YPEPL100k50.eps,width=4cm}
%\hspace{1cm}
%\epsfig{file=Bilder/kinetic/Spielmodell/YPEPL100k100nu03.eps,width=6.5cm}
\caption{Normalized real part of the admittance of the PEP with different maximal expansion indices in velocity space: $\kappa_{\rm max}=\lambda_{\rm max}=2$ (dashed) and $\kappa_{\rm max}=4$, $\lambda_{\rm max}=2$ (bold).}
\label{PEPL100nu02A}
\end{figure}   

In fig. \ref{PEPL100nu02A} the real part of the admittance is depicted for the maximum expansion indices in velocity space $\kappa_{\rm max}=\lambda_{\rm max}=2$ (dashed). On can observe a clear resonance peak below the plasma frequency at approximately $\omega_r=0.3297\,\omega_{p}$. Due to the geometry of the PEP this resonance can be interpreted as the series resonance of the probe. From the well known formula \cite{godyak1986,annaratone1995}
\begin{equation}
\omega_{\rm PSR}=\sqrt{\frac{2\delta}{L}}\,\omega_{\rm pe}\label{PSRparallel}
\end{equation}
one can determine a sheath thickness of about $\delta=5.435\,\lambda_D$. This sheath thickness represents a typical value in low-pressure plasmas in planar geometry and demonstrates the physically reasonable choice of the parameters $A$, $a$, and $L$. 

\begin{figure}[h!]
\centering
\psfrag{Y}{{\tiny $\Re\{Y\}\ [a.u.]$}}
\psfrag{o}{{\small $\frac{\omega}{\omega_p}$}}
\epsfig{file=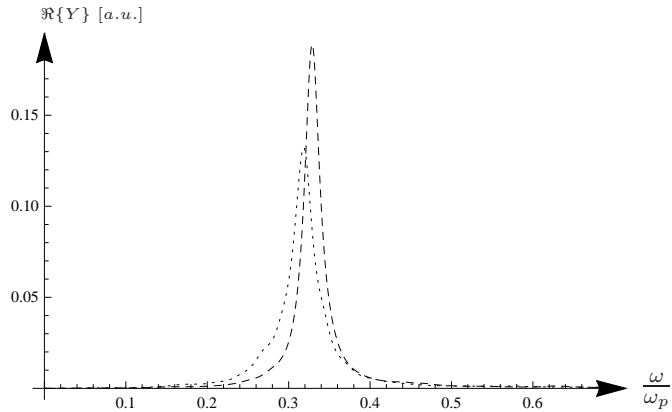,width=8.5cm}
%\hspace{8mm}
%\epsfig{file=Bilder/kinetic/Spielmodell/YPEPL100k50.eps,width=4cm}
%\hspace{1cm}
%\epsfig{file=Bilder/kinetic/Spielmodell/YPEPL100k100nu03.eps,width=6.5cm}
\caption{Normalized real part of the admittance of the PEP with different maximal expansion indices in velocity space: $\kappa_{\rm max}=\lambda_{\rm max}=2$ (dashed) and $\kappa_{\rm max}=4$, $\lambda_{\rm max}=3$ (dotted).}
\label{PEPL100nu02B}
\end{figure}  

To show the influence of kinetic effects on the damping of the resonance peak we increase the maximum expansion indices in velocity space. In fig. \ref{PEPL100nu02A} only the maximum expansion index of the velocity's absolute value is increased to $\kappa_{\rm max}=4$ (bold). The resulting resonance peak is stronger damped than the first peak and slightly shifted to a lower resonance frequency of about $\omega_r=0.328\,\omega_p$. A much stronger damping can be observed by an additional increase of the maximum expansion index of the projection angle $\lambda_{\rm max}=3$ (dotted) as depicted in fig. \ref{PEPL100nu02B}. This additional damping leads also to a meaningful shift of the resonance frequency to $\omega_r=0.319\,\omega_{p}$, which can be explained by an additional kinetic term in the complete collision frequency $\nu=\nu_0+\nu_{\rm kin}$. Hence, a better resolution within the velocity space shows a stronger damping and demonstrates clearly the influence of kinetic effects on the damping of the resonance in the spectrum of APRS. %The shift to the lower resonance frequency $. 

\section{Conclusion}

Based on the result that the admittance of a probe-plasma system in a kinetic description of APRS in arbitrary geometry is given by the resolvent of the dynamical operator $\op T_V + \op T_S$, we derived the expanded admittance for the PEP. Therefore, a complete basis in the particular geometry was chosen to determine the matrix representation of the dynamical operator. The truncation of the expansion leads to the approximated admittance which allows to analyse the influence of kinetic effects on the spectrum and in particular on the damping.  

To demonstrate that kinetic effects have a meaningful influence on the damping of an APRS probe in a low-pressure plasma, we compared three different spectra. They differ in their maximum expansion indices in velocity space, but they are computed with the same maximum expansion index in physical space. A large expansion index in physical space is necessary to ensure converged spectra in all cases. 

An increase only of the maximum expansion index of the velocity's absolute value shows just a weak influence on the damping. Such an increase is connected to a better resolution of the electron's kinetic energy. This energy has an influence on the kinetic free energy, which escapes through the interface $\set F$ in the geometry of a realistic probe design, but it is always limited due to elastic collisions. In case of the PEP the interface $\set F$ lies in the outer region of the electrodes, which means, that the spectra do not fully contain this loss mechanism. Furthermore, the collision frequency was assumed to be constant. A velocity dependent collision frequency will have a stronger influence. 

An additional increase of the maximum expansion index of the projection angle leads to a stronger influence on the damping. It is caused by elastic collisions which drive the perturbed distribution function to an isotropic one. This means to lose information about the velocity direction. Information loss is equal to an increase of the kinetic entropy and thus, a decrease of the kinetic free energy. 

In summary we have shown that kinetic effects have an influence on the loss of kinetic free energy which is observed by the probe as damping. On one hand it is caused by the increase of the kinetic entropy due to elastic collisions and on the other hand by the escaped free energy to an unobservable distance to the probe. The latter is connected to the collisionless damping. 

As a next step the presented solution strategy of the kinetic model will be applied to a real probe design. This will lead to a correction of a fluiddynamically determined half-width of the resonance peak. The kinetically corrected half-width is dependent of the electron temperature and will allow for a simultaneous measurement of the electron density and temperature with an APRS probe.

\acknowledgments
The authors acknowledge support by the internal funding of Leuphana University, support by the Federal Ministry of Education and Research (BMBF) in frame of the project PluTO+, and support by the Deutsche Forschungsgemeinschaft (DFG) via Collaborative Research Center TR 87,\linebreak and the Ruhr University Research School. Gratitude is expressed to D.-B. Grys, M.~Lapke, M.~Oberberg, C.~Schulz, J.~Runkel, R.~Storch, T.~Styrnoll, S.~Wilczek, T.~Mussenbrock, P.~Awakowicz, T.~Musch, and I.~Rolfes, who are or were part of the MRP-Team at Ruhr University Bochum.

\pagebreak
%\clearpage

\end{document}